\documentclass[conference]{IEEEtran}
\makeatletter
\def\ps@headings{%
\def\@oddhead{\mbox{}\scriptsize\rightmark \hfil \thepage}%
\def\@evenhead{\scriptsize\thepage \hfil \leftmark\mbox{}}%
\def\@oddfoot{}%
\def\@evenfoot{}}
\makeatother
\usepackage[hyphens]{url}
\usepackage{soul}
\usepackage{amsfonts}
\usepackage{mathrsfs}
\usepackage{amssymb}
\usepackage{graphicx}
\usepackage{epsfig}
\usepackage{epstopdf}
\usepackage{psfrag}
\usepackage{amsmath}
\usepackage{array}
\usepackage{booktabs}
\usepackage{subfigure}
\usepackage{cite,color}
\usepackage{anysize}
\usepackage{algorithmic}
\usepackage{algorithm}
\usepackage{stmaryrd}
\usepackage{multirow}
\usepackage{times}
\usepackage[dvipsnames]{xcolor}
\usepackage{authblk}
\usepackage{hyperref}

\DeclareGraphicsExtensions{.eps,.pdf,.png,.jpg,.gif,.jpeg,.pstex}

\marginsize{13mm}{13mm}{13mm}{13mm}
\IEEEoverridecommandlockouts

\graphicspath{{figure/}}
\title{Experiments with mmWave Automotive Radar Test-bed}
\author[1]{Xiangyu Gao}
\author[1]{Guanbin Xing}
\author[1]{Sumit Roy}
\author[1,2]{Hui Liu}
\affil[1]{Department of Electrical and Computer Engineering, University of Washington}
\affil[2]{Silkwave Holdings}
\affil[ ]{\textit{\{xygao, gxing, sroy, huiliu\}@uw.edu}}

\begin{document}
\maketitle

\section{Abstract}
Millimeter-wave (mmW) radars are being increasingly integrated in commercial vehicles to support new Adaptive Driver Assisted Systems (ADAS) for its ability to provide high accuracy location, velocity, and angle estimates of objects, largely independent of environmental conditions. Such radar sensors not only perform basic functions such as detection and ranging/angular localization, but also provide critical inputs for environmental perception via object recognition and classification. To explore radar-based ADAS applications, we have assembled a lab-scale frequency modulated continuous wave (FMCW) radar test-bed (\href{https://depts.washington.edu/funlab/research}{\textit{https://depts.washington.edu/funlab/research}}) based on Texas Instrument's (TI) automotive chipset family. In this work, we describe the test-bed components and provide a summary of FMCW radar operational principles. To date, we have created a large raw radar dataset for various objects under controlled scenarios. Thereafter, we apply some radar imaging algorithms to the collected dataset, and present some preliminary results that validate its capabilities in terms of object recognition. Our code is available at \href{https://github.com/Xiangyu-Gao/mmWave-radar-signal-processing-and-microDoppler-classification}{\textit{https://github.com/Xiangyu-Gao/mmWave-radar-signal-processing-and-microDoppler-classification}}.

\begin{figure}[h]
\centering
\subfigure[]{\includegraphics[width=1.65in]{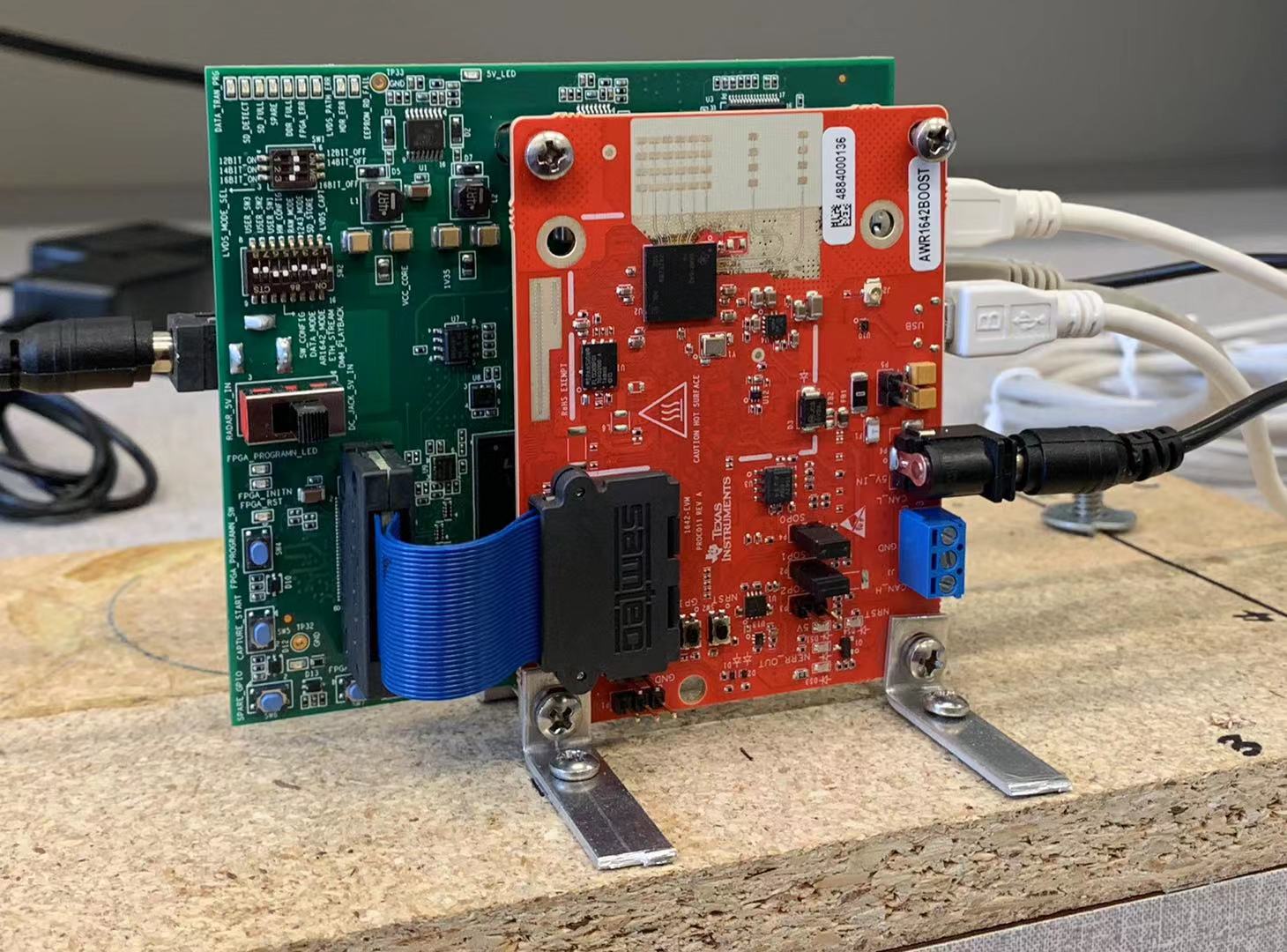}}
\subfigure[]{\includegraphics[width=1.7in]{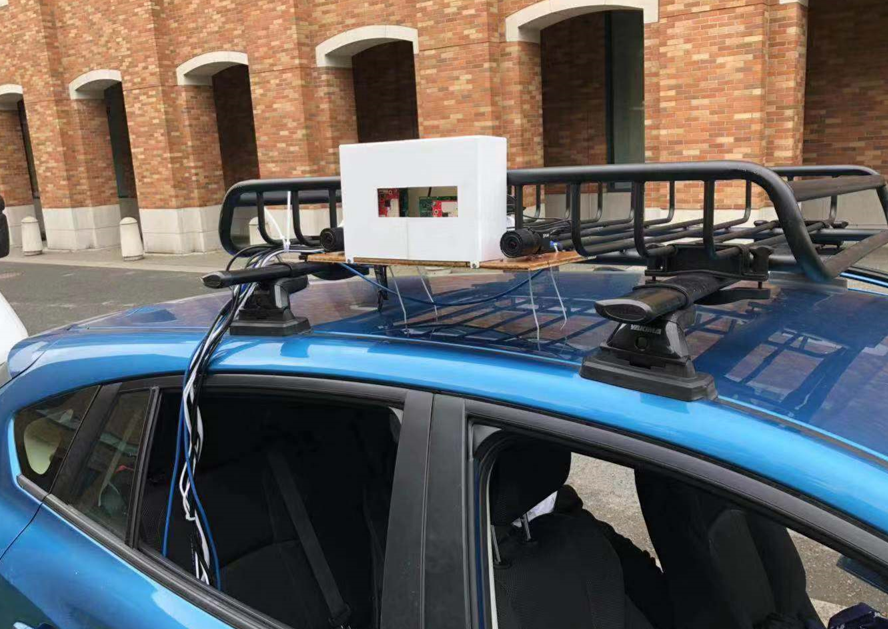}}
 \vspace{-3mm}
\caption{(a) FMCW radar test-bed (red board: AWR1642 BOOST; green board: DCA1000 EVM)  (b) Vehicle mounted platform for dataset collection}
  \label{test-bed}
\end{figure}

\section{Introduction}
 Over the years, advances in 77GHz RF design with integrated digital CMOS and packaging have enabled low-cost radar-on-chip and antenna-on-chip systems \cite{8828025}. As a result, several vehicular radar vendors are refining their radar chipset solutions for the automotive segment. TI's state-of-art 77GHz FMCW radar chips and corresponding evaluation boards - AWR1443, AWR1642, and AWR1843 - are built with the low-power 45-nm RF CMOS process and enable unprecedented levels of integration in an extremely small form factor \cite{ti1642datasheet}. Uhnder has also recently unveiled a new, {\em all-digital} phase modulated continuous wave (PMCW) radar chip that uses the 28nm RF CMOS process and is capable of synthesizing multiple input multiple output (MIMO) radar capability with 192 virtual receivers, thereby obtaining a finer angular resolution \cite{8662386}. However, compared to FMCW radars, PMCW radars shift the modulation complexity/precision to the high-speed data-converters and the DSP. Overall, continual progress in radar chip designs is expected to enable further novel on-platform integration and, consequently lead to enhanced performance in support of ADAS elements such as adaptive cruise control, auto emergency braking, and lane change assistance \cite{8828025}. 
 
 The above applications fundamentally rely on advanced radar imaging, detection, clustering, tracking, and classification algorithms. Significant research in the context of automotive radar classification has demonstrated its feasibility as a good alternative when optical sensors fail to provide adequate performance. \cite{6042174} reported that with handcrafted  feature extraction from range and Doppler profile, over $90\%$ accuracy can be achieved when using the support vector machine (SVM) algorithm to distinguish cars and pedestrians. Other studies used the short-time Fourier transform (STFT) intensity (heatmap) as the input for object classification. \cite{8468324} used different deep learning methods to extract micro-Doppler patterns in the STFT heatmap, with up to $93\%$ recognition accuracy when evaluated for three class discrimination: car, pedestrian and cyclist. However, the data of \cite{8468324} was obtained solely from single input, single output (SISO) radar where only one range bin was used to generate  STFT heatmap. Compared to MIMO radar, SISO contains little information about the shape of extended objects, that can greatly improve object recognition and classification. Also, the evaluation of \cite{8468324} didn't take the detection error into account and hence the accuracy metric would not demonstrate its ability to deal with missing detection and false alarm.
 
 In our work, we first review basic operation principles and imaging algorithm for FMCW radar. We propose a new framework to pre-process the raw radar data and improve object classification by combining both the spatial and temporal information. In our method, the radar data-cube is cropped from the Range-Angle (RA) heatmap based on the constant false alarm rate (CFAR) \cite{doi:10.1036/0071444742} detection algorithm (Fig. \ref{rawpre}) and then processed to extract target micro-Doppler signature (Fig. \ref{stftpre}), which is the input to the VGG16 deep learning classifier \cite{1409.1556}. This algorithm is implemented on a huge dataset and the performance compared with a baseline is evaluated with two metrics - {\em precision, and recall} - to quantify the improvements from our approach. 
 
 The rest of the paper is organized as follows. Section 3 introduces our automotive radar test-bed setup. Section 4 describes the basic measurement theory for FMCW radar and our signal processing workflow for radar imaging. Section 5 describes our dataset collection effort and shows some baseline imaging examples. Section 6 describes the implementation and performance of our proposed radar object classification algorithm. Section 7 concludes the paper.

\section{Radar Test-bed}
\begin{figure}[h]
\centering
\includegraphics[width=3in]{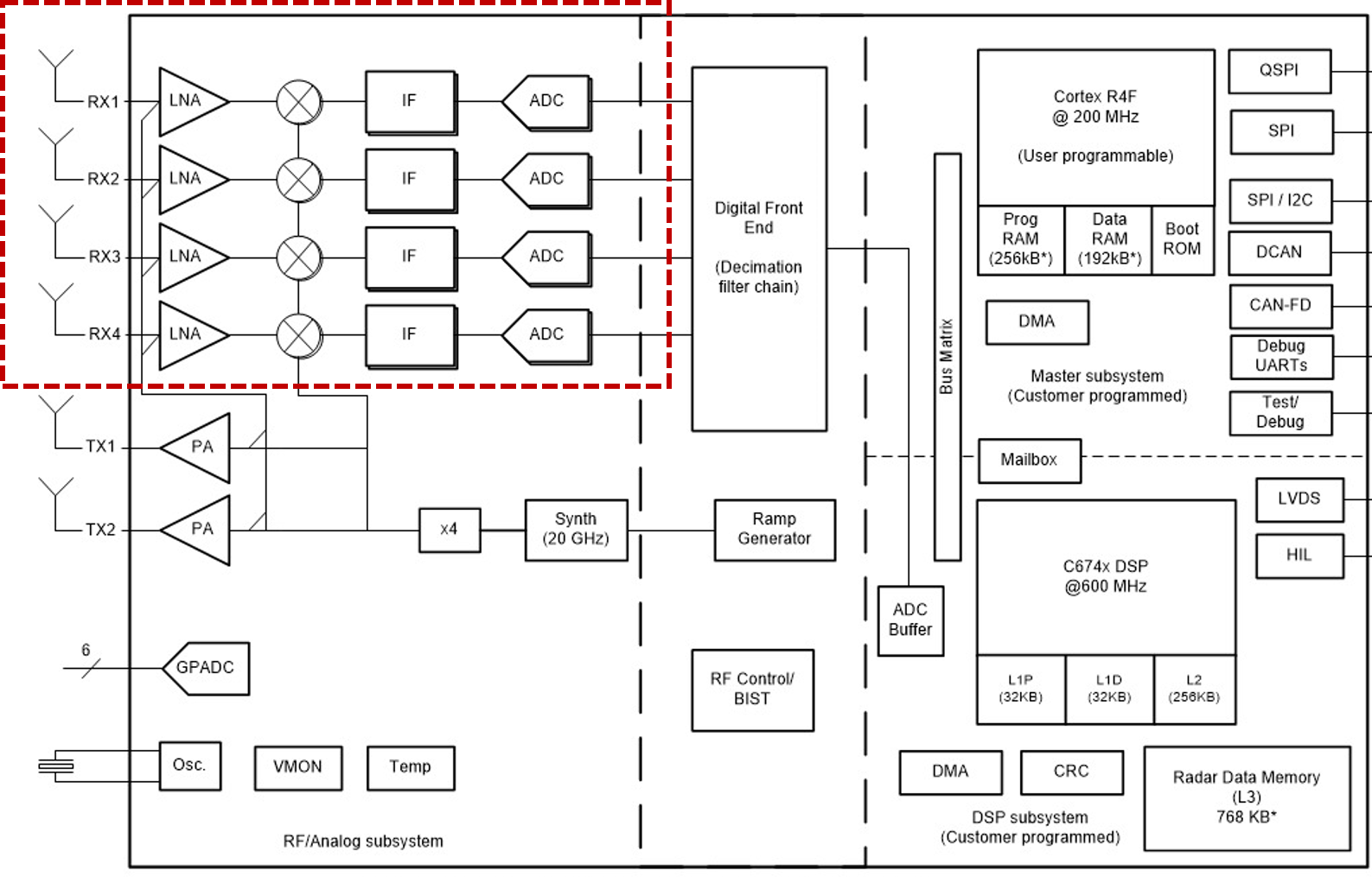}
  
\caption{TI AWR1642 chip diagram, where the red dash rectangle highlights the  receiver chain \cite{ti1642datasheet}}
  \label{ti1642}
\end{figure}

Our automotive radar test-bed (Fig. 1a) is composed of two TI evaluation boards: AWR1642 BOOST and DCA100 EVM. As shown in Fig. \ref{ti1642}, the AWR1642 chipset is an integrated FMCW radar sensor that enables a monolithic implementation of a 2TX, 4RX system with built-in phase lock loop (PLL) and analog to digital converters (ADC) \cite{ti1642datasheet}. The RF design makes it capable of operation in the 76-77 or 77-81GHz band with 12.5dBm TX power and 15dB RX noise figure. It also integrates the C674x-based DSP subsystem and ARM R4F-based processor subsystem, which are responsible for radar signal processing and radio configuration control, respectively.
DCA1000 EVM is a capture board for streaming the ADC data from AWR1642 board to a local computer over Ethernet.

\section{Basic Measurement Theory}

\subsection{Range Measurement}
\begin{figure}
\centering
\includegraphics[width=3in]{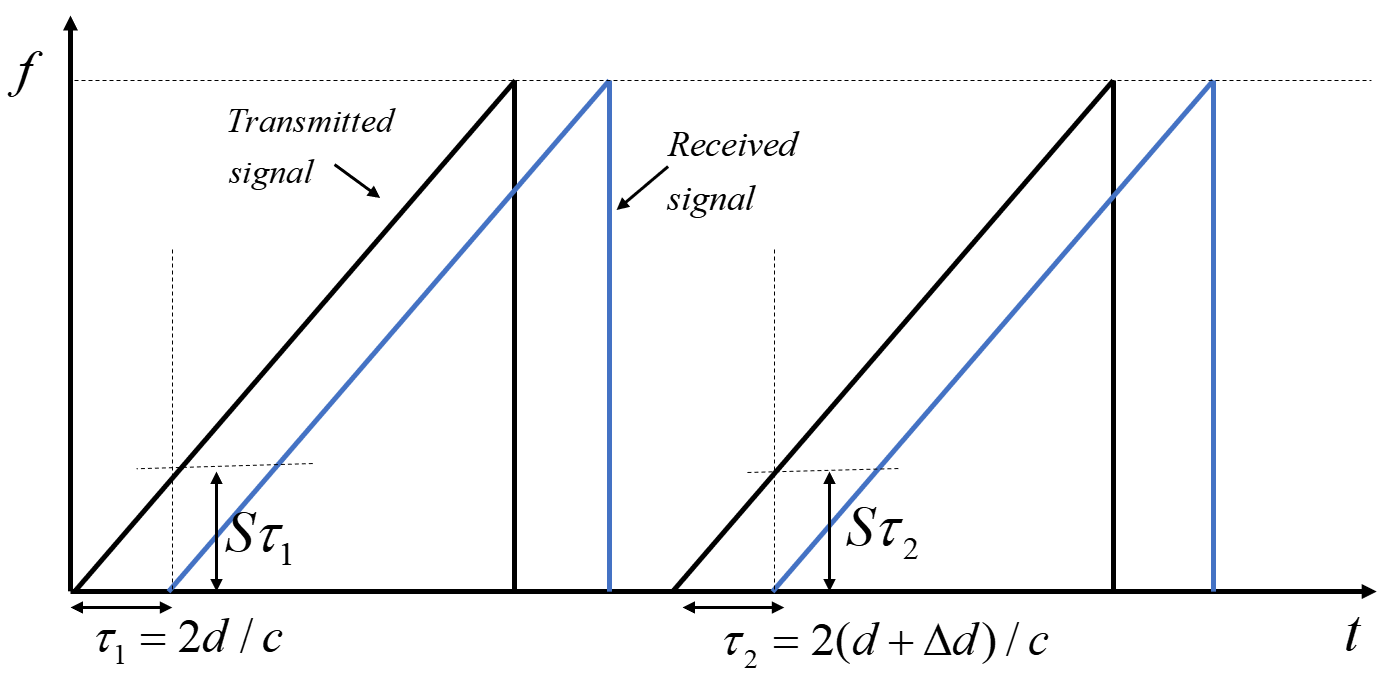}
  
\caption{Range and Velocity Measurement}
  \label{fig1}
\end{figure}

As shown in Fig.\ref{fig1}, an FMCW radar transmits a sequence of chirp signals (called a frame) and then mixes the receive echo with the local reference (transmitted signal) to yield a resulting beat signal at a frequency $f_b=\frac{S2d}{c}$ in the intermediate frequency (IF) band (shown in Fig. \ref{fig1}), where $S$ is the slope of chirp signal, $d$ is the distance to the object, and $c$ is speed of light. To estimate the beat frequency, it is common to use a fast Fourier transform ({\em Range FFT}) to convert the time domain IF signal into the frequency domain, and the resulting spectrum has separate peaks for resolved objects.

The resolution of FFT-based range estimation is determined by the swept RF bandwidth $B$ of the FMCW system \cite{doi:10.1036/0071444742}, given by the well-known result $R_{res}=\frac{c}{2B}$. In our following experiments, the FMCW signal is configured with 670 MHz swept bandwidth, and the expected range resolution is 0.23m.

\subsection{Velocity Measurement}

 The object motion $\Delta d$ (shown in Fig. \ref{fig1}) relative to the radar causes a beat frequency shift $\Delta f_b = \frac{2S\Delta d }{c}$  on the receive signal as well as a phase shift $\Delta \phi_v = 2\pi f_c\frac{2\Delta d}{c}=\frac{4\pi v T_c}{\lambda}$, where $f_c$ is the center frequency, $v$ is the object velocity, $T_c$ is the chirp duration, and $\lambda$ is the wavelength. Compared to the beat frequency shift, the phase shift of mmW signal is more sensitive to the object movement. Hence, it is common to execute a fast Fourier transform ({\em Velocity FFT}) across the chirps to estimate the phase shift and then transform it to velocity.

The velocity resolution of this method is given by: $V_{res}=\frac{\lambda}{2T_f}=\frac{\lambda}{2LT_c}$ \cite{iovescu2017fundamentals}, where $L$ is the number of chirps in one frame, and $T_f$ is the frame period. The expected velocity resolution is 0.065m/s for our dataset given the configuration $L=255$, and $T_c=120us$.

\subsection{Angle Measurement}
Angle (azimuth) estimation is conducted via processing the signal at a received phased array composed of multiple elements.
\subsubsection{FFT Algorithm}
The return from an object at a sufficient distance, located at an azimuth angle $\theta$ results in a phase difference of $\Delta \phi_\theta = \frac{2\pi h\sin{\theta}}{\lambda}$ between any adjacent pair of RX antennas, where $h$ is the separation between receive antenna pair. Hence a fast Fourier transform in the spatial dimension across RX elements ({\em Angle FFT}) resolves objects according to their arrival angles in azimuth. The angle resolution for FFT processing is known to be  $\theta_{res}=\frac{\lambda}{N_{RX}h\cos{\theta}}$ \cite{iovescu2017fundamentals}, where $N_{RX}$ is the number of receive antennas. The expected angle resolution of our MIMO radar is about $15^{\circ}$ at azimuth $\theta = 10^{\circ}$ given $N_{RX}=8$, $h=\frac{\lambda}{2}$.

\subsubsection{Multiple Signal Classification (MUSIC) Algorithm}

As is well-known, enhanced (compared to Angle FFT processing) angular resolution can be potentially obtained via the class of high-resolution Angle-of-Arrival (AoA) estimation algorithms, a prominent representative being the Multiple Signal Identification and Classification (MUSIC) algorithm. MUSIC belongs to the class of {\em eigen-decomposition} based AoA estimators that construct the $(N-K)$-dim. noise subspace $G$ and $K$-dim. signal subspace $Z$ (where $K < N$ denostes the number of targets) from the covariance matrix of received signals $y(t)$\cite{1396448}. The MUSIC estimate of the $K$ DoAs ${\{\theta_k\}}_{k=1}^K$ are given by the locations of $K$ minimas of the function: $f(\theta) = {a^*(\theta)(I-ZZ
^*)a(\theta)}$, where $a(\theta)$ denotes the steering vector corresponding the azimuth look direction $\theta$. 

\subsection{Signal Processing Workflow and Radar Imaging}
\subsubsection{Radar signal Processing Work Flow} As shown in 
Fig. \ref{workf}, the Range FFT is employed on time domain ADC samples to obtain the range estimate. Then the Velocity FFT across the chirps of one frame is executed for the velocity estimation. Also, short-time Fourier transforms on the Range FFT results generate the STFT heatmap that can track the object velocity. To produce the RA heatmap, we detect the maximum Doppler peak at each range bin, and then perform Angle FFT at the detected Doppler bins. All Angle FFT results are concatenated together to form the final RA heatmap. This procedure is named as 3-dimension FFT (3DFFT) as we employ three fast Fourier transforms (Range FFT, Velocity FFT, and Angle FFT). We can also obtain the 3D point cloud by implementing Angle FFT on the CFAR detections of range-velocity bins.

\subsubsection{Range-Angle (RA) Heatmap}

RA heatmap is a two-dimensional (azimuth angle, range) image that shows the spatial pattern of the received echo, see Fig. \ref{fig_imag_examp} (middle column). The brighter colors in RA heatmap represents greater reflection amplitude at that (range, angle) location. 

\subsubsection{Short-Time Fourier Transform (STFT) Heatmap}

STFT heatmap is a time-frequency image that plots the spectra as a function of time, see Fig. \ref{fig_imag_examp} (right column). Compared to the usual Velocity FFT measurement mentioned above, STFTs are computed as follows: divide a longer duration time signal into shorter segments of equal length and compute the Fourier transform separately on each shorter segment. The sequence of resulting Fourier spectra can be used to estimate and track object velocity and display specific micro-Doppler signatures corresponding to different objects.

\begin{figure}[h]
\vspace{-3.5mm}
\centering
\includegraphics[width=3.7in]{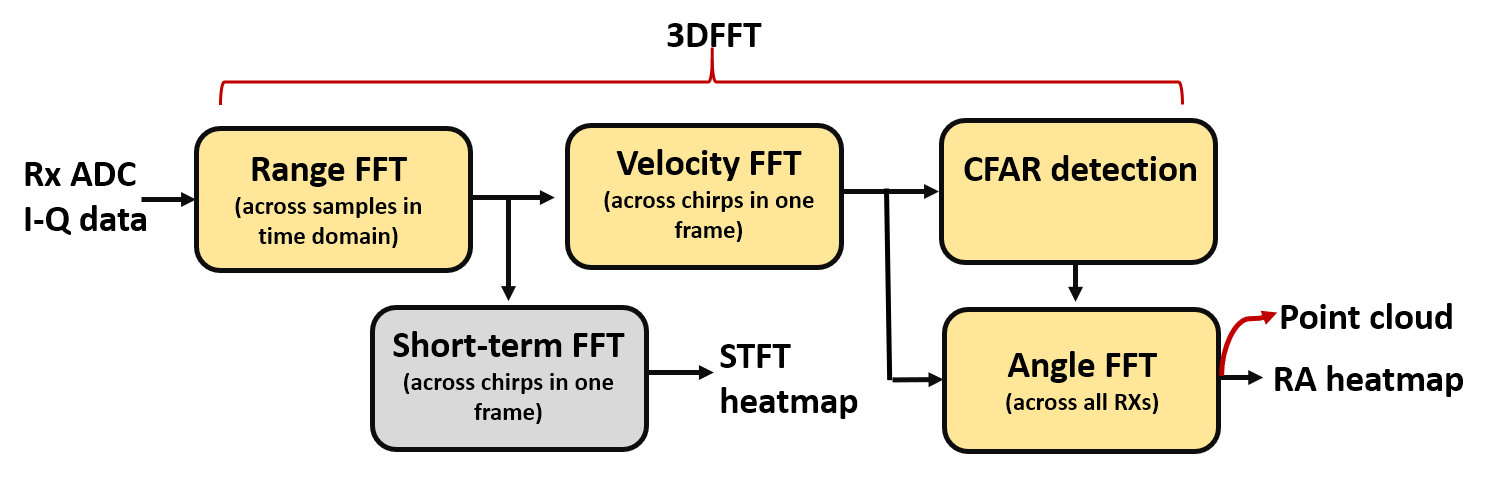}
  
\caption{Basic radar signal processing chain}
  \label{workf}
\end{figure}

\section{Dataset Collection and Radar Imaging Examples}

\subsection{Radar Dataset Collection}

A large radar dataset (raw I-Q samples post demodulation at receiver) for various objects have been collected for multiple scenarios -  parking lot, campus road, city road, freeway by a vehicle mounted platform that is driven (see Fig. \ref{test-bed}b). In particular, significant effort was placed in collecting data for situations where cameras are largely ineffective, i.e. under challenging light conditions. The collected data is used both for validation/calibration (training) and radar imaging (testing) for object discrimination for primarily three classes: pedestrian, cyclist and cars. The radar configurations used for the data collection are shown in Table \ref{paracon} and the data has been labeled with object location and class. In Fig. \ref{fig_imag_examp}, several typical examples from our dataset and corresponding radar images are demonstrated\footnote{We intend to release a portion of our dataset which includes the raw radar data and corresponding metadata (labels) for 2 classes of objects (pedestrian and car). Please contact us if you are interest in our Automotive Radar work and the data. (xygao@uw.edu).}. 

\begin{table}[ht]
\begin{center}
\caption{Configurations of FMCW signal and test-bed}
\begin{tabular}{cc}
\toprule  
Parameter & Configuration\\
\midrule  
Start Frequency & 77 GHz\\
Sweep Bandwidth & 670 MHz\\
Sweep slope & 21 MHz/us \\
Frame rate & 30 fps \\
Sampling frequency & 4000 ksps\\
Number of chirps in one frame & 255\\
Number of samples of one chirp & 128 \\
Number of transmitters, receivers & 2, 4 \\
\bottomrule 
\label{paracon}
\end{tabular}
\end{center}
\vspace{-6mm}
\end{table}

\subsection{Data Validation}
\subsubsection{Maximum range}

To verify the maximum range of the TI platform, we compared our experimental results with predictions based on the radar range equation \cite{doi:10.1036/0071444742}. Ten measurements of the maximum detection range for pedestrian were conducted and the averaged result was found to be 24.2m for the cell average-CFAR (CA-CFAR) algorithm with the 2dB threshold.

In \cite{doi:10.1036/0071444742}, the radar range equation is given by

\begin{equation}
    \label{eq:6}
R_{max} = \sqrt[4]{\frac{P_t G_T G_R \lambda^2 \sigma}{(4\pi)^3 P_{min}}}
\end{equation}

where $P_t$ is the transmit power, $G$ is the Antenna Gain, $\lambda$ is the transmit wavelength, $\sigma$ is the target radar cross section, and $P_{min}$ is the minimum detectable power.

The minimum detectable power at receiver input is 
\begin{equation}
    \label{eq:7}
P_{RX,min} = kTB_RF(S/N)_{min}
\end{equation}
where $k$ is the Boltzmann's Constant, $T$ is the temperature in degrees Kelvin, $B_R$ is the Receiver Bandwidth\footnote{$B_R$ equals the IF bandwidth (same as LPF bandwidth) in the radar receive chain shown in the red dash rectangle of Fig. \ref{ti1642}}, $F$ is the receiver Noise Figure and $(S/N)_{min}$ is the minimum signal to noise ratio required at receiver input for reliable detection. The parameter values needed for our analysis were obtained from TI's AWR1642 datasheet\cite{ti1642datasheet}.\\

\begin{table}[ht]
\begin{center}
\caption{Maximum Range Calibration parameters}
\begin{tabular}{cccc}
\toprule  
Parameter & Value & Parameter & Value\\
\midrule  
$P_t$ & 12.5 dBm & k & $1.38 \times 10^{-23}$\\
$G_T$ & 1 & T & $290^\circ K$\\
$G_R$ & 34 dBi & $B_R$ & 4 MHz\\
$\lambda$ &  0.0038961 m & F & 15 dB\\
$\sigma$ & $1m^2$ \cite{6229134}& $ (S/N)_{min} $ & 2 dB\\
\bottomrule 
\label{paracon}
\end{tabular}
\end{center}
\vspace{-6mm}
\end{table}

Utilizing Eq. (\ref{eq:6}), (\ref{eq:7}), the calculated $P_{RX,min} \approx -91 dBm$ and corresponding maximum range is about 25m, which is very close to the experimental result.

\begin{figure}[h]
\centering
\subfigure[]{\includegraphics[width=1.17in]{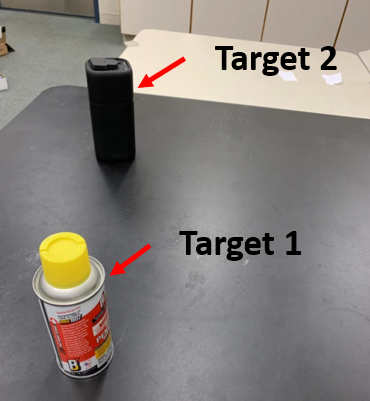}}
\subfigure[]{\includegraphics[width=1.67in]{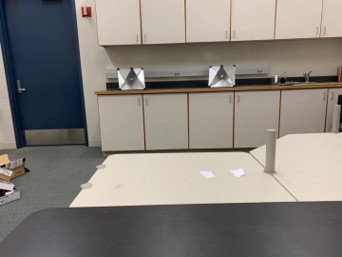}}
\vspace{-3.5mm}
\caption{Experiment setup: (a) two small static objects  (b) two corner reflectors}\label{exp_setup}
\end{figure}

\subsubsection{FFT-based Range resolution}
\begin{figure}[h]
\centering
\subfigure[]{\includegraphics[width=1.73in]{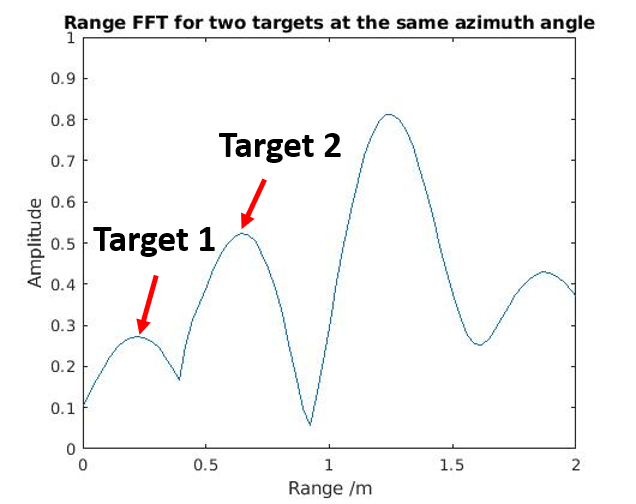}}
\subfigure[]{\includegraphics[width=1.73in]{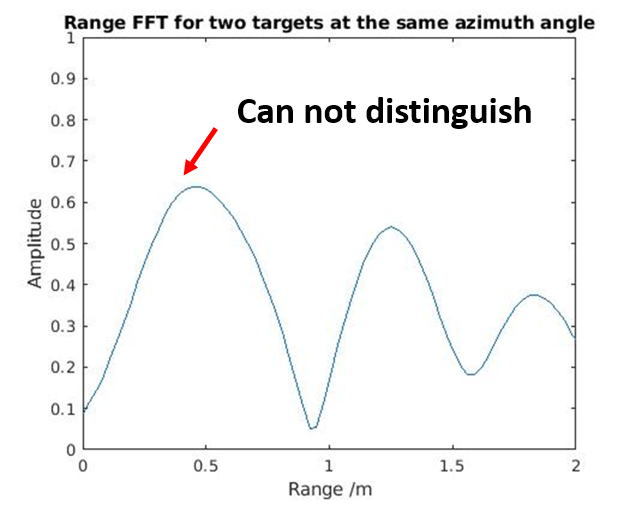}}
\vspace{-3mm}
  \caption{(a) FFT-based range estimation for two static targets with 0.3m range difference  (b) FFT-based range estimation for two static targets with 0.23m range difference}\label{fig_rng_cal}
\end{figure}

As shown in Fig. \ref{exp_setup}a, we placed two small targets at different locations to verify the minimum distinguishable range difference. When the range difference was 0.3m, two targets could be separated via Range FFT (Fig. \ref{fig_rng_cal}a). However, when the separation was reduced to 0.23m, two targets could not be resolved by Range FFT anymore (Fig. \ref{fig_rng_cal}b).

 Our experiment result shows that the best range resolution in the lab is 0.3m which is greater than the theoretical value 0.23m by a little bit. One possible reason for the degradation is the spectrum widens as the source is not an ideal point target.  

\subsubsection{DoA estimation algorithm comparison}

\begin{figure}[h]
\vspace{-3.5mm}
\centering
\subfigure[]{\includegraphics[width=1.71in]{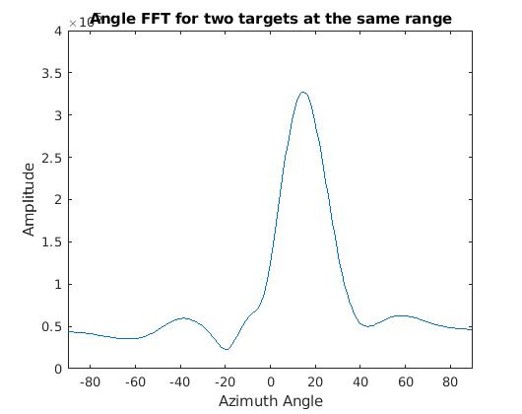}}
\subfigure[]{\includegraphics[width=1.78in]{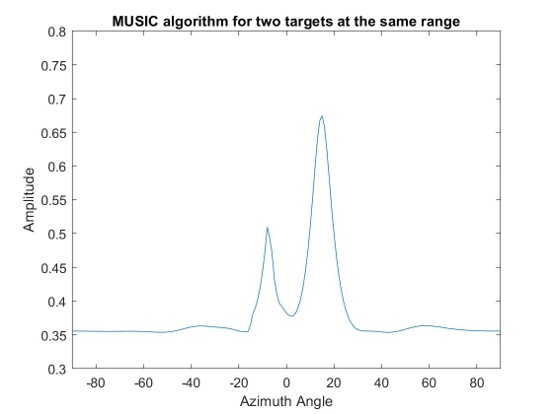}}
  \vspace{-3mm}
  \caption{(a) FFT-based DoA estimation result for two corner reflectors with $20^{\circ}$ azimuth difference  (b) MUSIC-based DoA estimation result for the same scenario as (a)}
  \label{fig_music_agl}
\end{figure}

To compare the performance of FFT-based and MUSIC-based DoA estimation algorithm, we placed two corner reflectors on the table with a mean value of $20^{\circ}$ azimuth difference (Fig. \ref{exp_setup}b). From Fig. \ref{fig_music_agl}a, FFT-based DoA estimation generated a single tone for two corner reflectors, while the MUSIC-based DoA estimation (Fig. \ref{fig_music_agl}b) could separate two tones which represent the azimuth angles of two corner reflectors.

\subsection{Radar imaging examples}

We present three examples of radar imaging corresponding to (i) moving pedestrian,  (ii) moving car, and (iii) moving cyclist.  For each scenario, we compare the ground truth camera image with the radar RA heatmap. As they are all moving objects, we also plot the STFT heatmap to demonstrate the ability of estimating and tracking object velocity.

\begin{figure}[ht]
    \vspace{-3.5mm}
    \hspace{-1em}
%  \centering
  \includegraphics[width=3.9in, trim=1 1 1 1,clip]{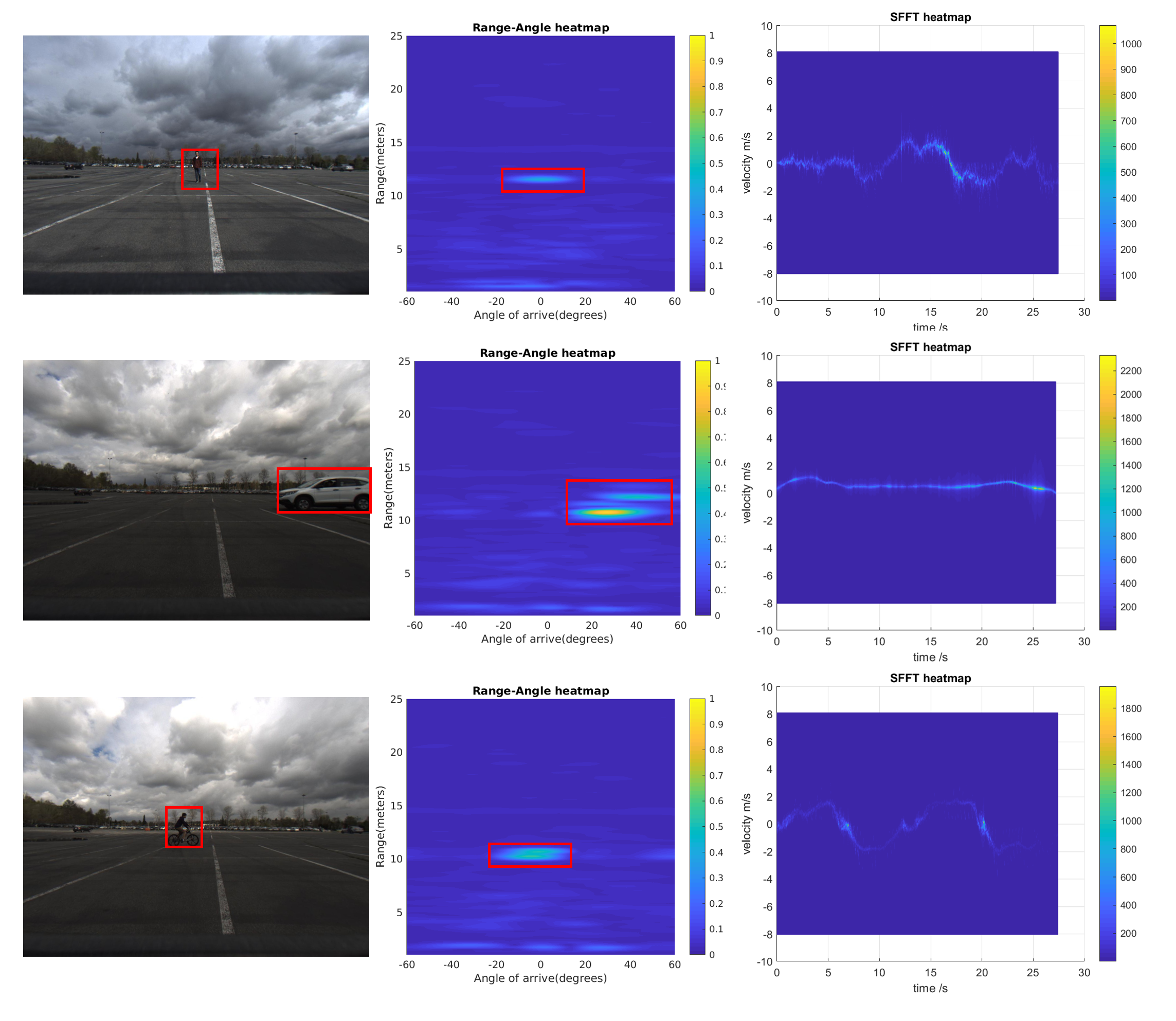}
  \vspace{-3mm}
  \caption{There are three rows that correspond to single moving pedestrian, single moving car and single moving cyclist respectively. For each row, there are camera image (left), RA heatmap (center), and STFT heatmap (right)}
  \label{fig_imag_examp}
\end{figure}

Fig. \ref{fig_imag_examp}a shows a pedestrian walking in the parking lot with normal speed. The generated RA heatmap (center column) illustrates the position (12m, $0^{\circ}$) of pedestrian at one time slot. The STFT heatmap (right column) displays specific micro-Doppler signature of a human gait which combines the Doppler shift of moving body and the micro-Doppler shifts of swinging arms and legs \cite{7348890}. Zooming in part of the STFT heatmap, we observe that the body$^\prime$s Doppler shift appears as the main velocity trajectory while the micro-Doppler shifts corresponding to arm and leg movements appear as vibrations in velocity with a periodic component. 

Fig. \ref{fig_imag_examp}b shows a slowly moving car in the parking lot. As cars have greater radar cross section (RCS) and bigger size, RA heatmap shows a larger group of reflections points that is distinct from the heatmap for a pedestrian. From the STFT heatmap, cars have smoother micro-Doppler signature as there are fewer moving parts in the car.

The RA heatmap and STFT heatmap of a moving cyclist is shown in Fig. \ref{fig_imag_examp}c. The RA reflection pattern of a cyclist is composed of the reflection from the bicycle and the reflection from a moving person. Similarly, the micro-Doppler signature of a cyclist is actually the combination of the movement of bicycle and the movement of the rider.

\section{Radar Object Classification}

The collected data was used to conduct radar classification testing for three classes: pedestrian, car, and cyclist. For our current test-bed, the high chirp rate of one frame yields enough temporal information revealing the movement pattern of a specific object. At the same time, the MIMO configuration provides the spatial information (e.g. shape) of the object. Hence, we propose a two-stages radar object classification framework which includes a CFAR detector and a deep learning classifier that automatically extracts features from the temporal and spatial information. We name the framework CDMC (CFAR detector and micro-Doppler classifier). To compare the performance of CDMC algorithm, we also design a simple decision tree baseline with handcrafted features.

\subsection{Proposed CDMC Framework}

\begin{figure}[h]
\vspace{-3.5mm}
\centering
\includegraphics[width=3.5in]{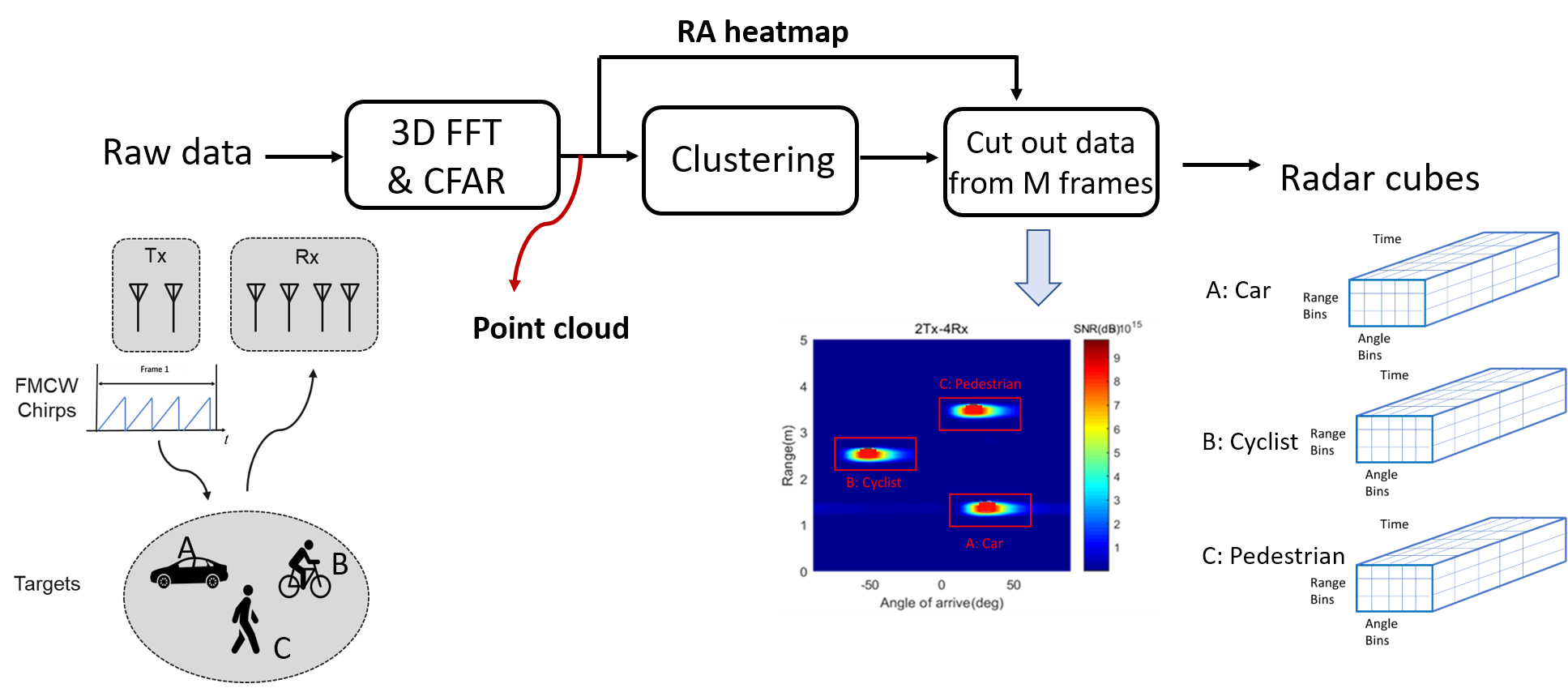}
  
\caption{Raw data pre-processing}
  \label{rawpre}
\end{figure}

\begin{figure}[h]
\vspace{-3.5mm}
\centering
\includegraphics[width=3.8in]{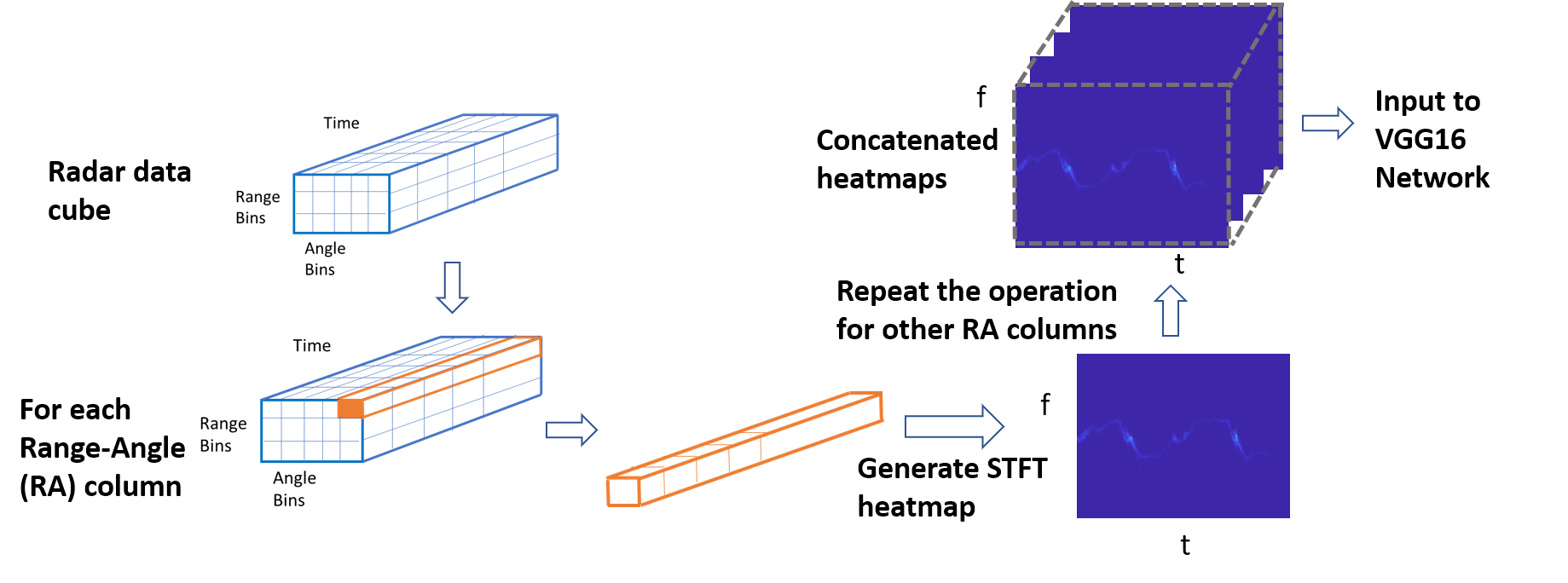}
  
\caption{STFT processing for radar data cube}
  \label{stftpre}
\end{figure}

The proposed CDMC framework has two stages: detection and classification. In the detection stage, we implement raw radar data processing (Fig. \ref{rawpre}) to acquire object location and the corresponding radar data cube. In the classification stage, we conduct STFT processing (Fig. \ref{stftpre}) on the obtained data cube to generate concatenated STFT heatmaps which serve as the input to the VGG16 deep learning classifier \cite{1409.1556}. 

In the raw data pre-processing part, we conduct the highlighted procedures (3DFFT and CFAR) shown in Fig. \ref{workf} to produce the RA heatmap and detection point cloud. Then we implement the DBSCAN \cite{Ester:1996:DAD:3001460.3001507} clustering algorithm to group together the point cloud and get the object center location. We set a fixed size bounding box for each detected object on the RA heatmap. The bounding box size should be large enough to cover most of the objects. We then take the radar data in the bounding box out of the $M$-frames RA heatmaps to form the radar data cube, where $M$ is the ``length'' of temporal information we aim to focus on. The reason why we use $M$-frames RA heatmaps instead of one is the temporal information, or object movement pattern, contained in consecutive $M$ frames can serve as good feature for classifying different objects.

As shown in Fig. \ref{stftpre}, the STFT processing part takes the radar cube as input which is the output of Fig \ref{rawpre}. We divide the radar data cube into several range-angle columns (i.e., orange column in Fig. \ref{stftpre}) which contain ``equal-length'' temporal information. Then we implement the STFT algorithm on one range-angle column to generate a time-frequency heatmap. By repeating this operation for other range-angle columns, we concatenate all generated STFT heatmaps together along the third dimension to form the input to the VGG16 classifier.

% \begin{figure}[h]
% \centering
% \includegraphics[width=3in]{vgg16.PNG}
  
% \caption{VGG16 deep learning classifier}
%   \label{vgg}
% \end{figure}

\subsection{Decision Tree Baseline}
For purposes of sanity check, a simple decision tree (DT) algorithm is used as the baseline. DT is a non-machine learning method and the parameter selection is based on empirical values. The input to DT is the point cloud obtained by 3DFFT and CFAR algorithm shown in Fig. \ref{rawpre}. We used two criterion to determine the object class given the point cloud information: (i) Is the size of the point cloud in range less than threshold $p$? If yes, we classify the object as a pedestrian. If no, we move on to the next question: (ii) Is the ratio of the object reflection amplitude to distance square greater than threshold $q$? If yes, we classify the object as a car; otherwise, it is a cyclist. 

The logic behind the above is straightforward: a pedestrian occupies a smaller spatial region than a cyclist or car and we use this feature (as in the first criterion) to discriminate. A car has a similar reflection shape to a cyclist but greater RCS and hence we use the ratio of averaged amplitude to distance square (2nd criterion) to distinguish between a car and cyclist. 

\begin{figure}[h]
\vspace{-3.5mm}
\centering
\includegraphics[width=3.2in]{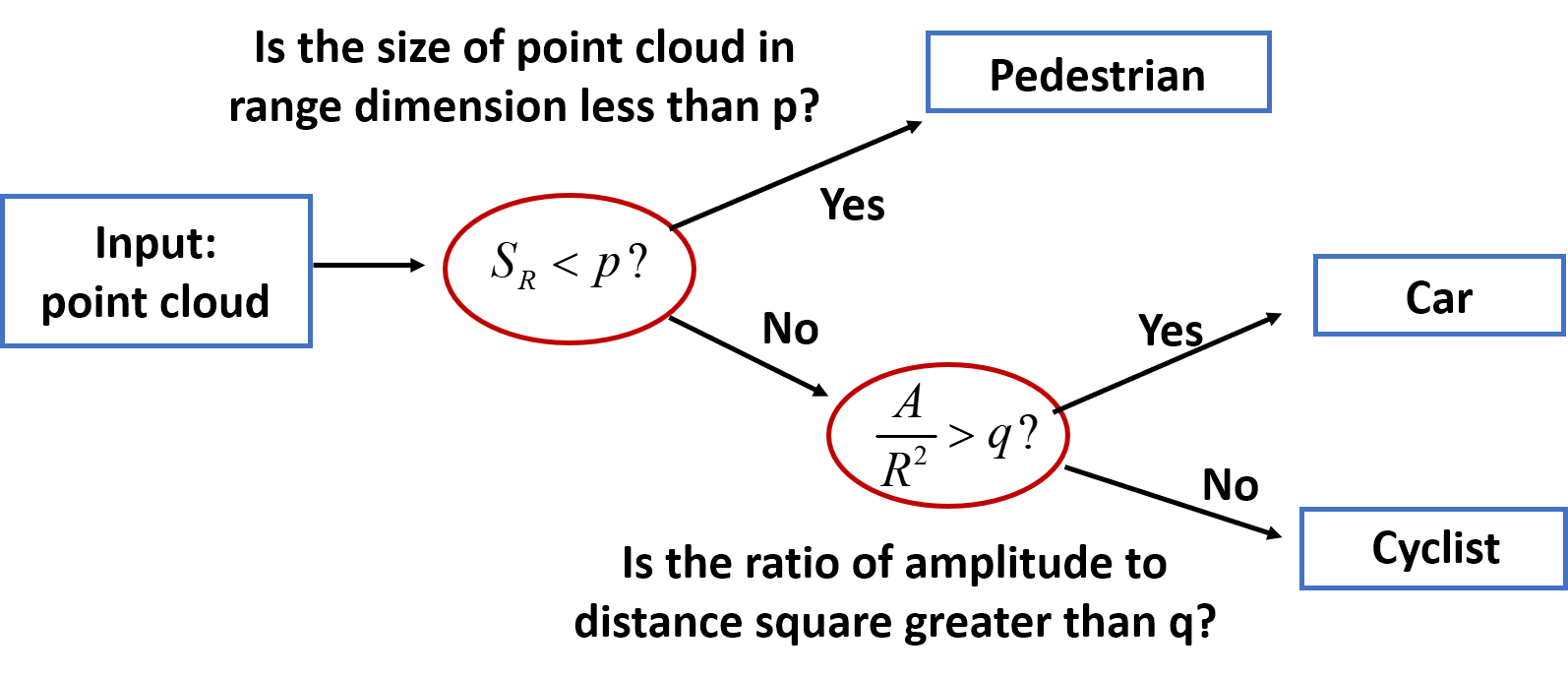}
  
\caption{Decision tree baseline, where $A$ is the averaged amplitude of point cloud, $R$ is the distance to radar,  $S_R$ is the point cloud size in range, i.e. the number of points in range, and $p,q$ are threshold scalars}
  \label{decision_t}
\end{figure}

\subsection{Experiment}

Data used in this experiment come from the dataset introduced in {\em Part A, Section 5}. A portion of the data was collected for a controlled scenario where a volunteer walked, rode the bike, and drove the car within the field of view of the current test-bed in a large empty parking lot. Other data were recorded in more real-world scenarios, e.g., keeping our mounted set-up parked next to actual road traffic. 

For the proposed CDMC algorithm, we implemented 128 points Range FFT, 256 points Velocity FFT, and 128 points Angle FFT to get the $128\times128$ size RA heatmaps at the raw data pre-processing stage (Fig. \ref{rawpre}). Then we determined the location of objects by using the CA-CFAR \cite{doi:10.1036/0071444742} detection algorithm and DBSCAN \cite{Ester:1996:DAD:3001460.3001507} clustering algorithm. We set an $11\times 5$ size bounding box for each detected object, where \textit{$11$} represents 11 range bins and \textit{$5$} represents 5 angle bins. We then took the radar data in the bounding box out of the 16-frames RA heatmaps to form the radar data cube.

With the data cube output from the first stage, we conducted the STFT processing (Fig. \ref{stftpre}) along the frame dimension to extract the micro-Doppler signature. The window size of STFT is 255; the overlap is 240; and the FFT size is 256. The output is $256\times 256 \times 55$ size concatenated STFT heatmaps, of which three dimensions represent the velocity, time, and different range-angle bins, respectively.

We generated about $1.2\times10^5$ concatenated STFT heatmaps in total, which were split into a training set, a validation set, and a testing set with the ratio 0.6/0.1/0.3. By feeding the training STFT heatmaps to the VGG16 classifier, we trained the model from scratch without using the pre-trained model. The total training epoch is 10; the batch size is 5; the learning rate is $10^{-4}$ for the first 5 epochs, and for the next 5 epochs, the learning rate is $10^{-5}$.

For DT algorithm, we tried different $p,q$ combinations on the validation set and chose $p=2$, $q=0.1$ according to the validation performance.

\subsection{Performance and Analysis}
We use two metrics to evaluate the performance of radar classification: precision and recall. They are defined as:
\begin{equation}
    \label{eq:9}
Precision = \frac{TP}{TP+FP} \, \quad 
Recall = \frac{TP}{TP+FN}
\end{equation}

where $TP$ is true positive which represents the number of objects correctly detected and classified, $FP$ is false positive which represents the number of background reflection points incorrectly classified as objects, and $FN$ is false negative which represents the number of undetected and/or incorrectly classified objects.

We test the DT baseline and CDMC algorithm in three scenarios: single object in an empty parking lot, multiple objects in an empty parking lot, and multiple objects in the crossroad. As shown in Table \ref{performance}, the proposed CDMC algorithm outperforms the DT baseline in both precision and recall. For the simple single object scenario, the DT baseline achieves 86.24\% precision and 57.53\% recall. The precision of DT is acceptable, as there is an 86.24\% chance for DT to correctly classify the detection as an object. However, the low recall represents the bad ability to correctly detect and classify the existing object in front of radar. In contrast, the recall of CDMC algorithm is 96.82\% which achieves a 68\% improvement over the DT baseline. At the same time, the precision increases to 92.03\%. 

Under more complicated scenarios, the precision and recall both decrease. For instance, in the  multiple pedestrian scenario, the recall of CDMC degrades to 89.58\% as there are some missing detection when several objects are close to each other or the reflection from the desired object is weak. On the other hand, the precision degrades considerably in the crossroad scenario because a large amount of reflection points from the background (e.g. building, tree) are detected and incorrectly classified as objects.
 
High recall algorithms are very important for autonomous driving. Let's think about two scenarios: (i) there is an object in front of a car but the car doesn't correctly detect and classify it, (ii) there is no object in front a car but the car makes a false alarm and classify it as an object. Apparently, the first scenario is more serious as the car will not respond to the potential danger. High-recall algorithms will eliminate the first dangerous scenario as much as possible. While the recall is guaranteed, we need to effectively reduce false alarm detection to improve precision.

\begin{table}[ht]
\begin{center}
\vspace{3.5mm}
\caption{The performance of radar classification algorithms}
\begin{tabular}{cccc}
\toprule  
Scenario & Method & Precision & Recall\\
\midrule  
\multirow{2}{*}{Single object, parking lot} & DT baseline & 86.24\% & 57.53\%\\
& CDMC algorithm & 92.03\% & 96.82\%\\

\midrule  
\multirow{2}{*}{Multiple pedestrians, parking lot} & DT baseline & 85.28\% & 78.01\%\\
& CDMC algorithm & 92.30\% & 89.58\%\\

\midrule  
\multirow{2}{*}{Multiple objects, crossroad} & DT baseline & 35.77\% & 44.84\%\\
& CDMC algorithm & 40.66\% & 80.68\%\\
\bottomrule 
\label{performance}
\end{tabular}
\end{center}
\vspace{-6mm}
\end{table}

\section{Conclusion and Future Work}
In this work, we described the mmW radar test-bed and provided an overview of the radar measurement theory. Thereafter, radar imaging and classification algorithms were applied to the collected dataset and results were presented. In our future work, we plan to optimize radar classification algorithm to improve precision performance.

\bibliography{bibtex}
\bibliographystyle{IEEEtran}

\end{document}